\documentclass[%
    aip,
    jcp,
    amsmath,
    amssymb,
    twocolumn,
    superscriptaddress,
    reprint]{revtex4-1}
\usepackage{dcolumn}
\usepackage{graphicx} 
\usepackage{bm}
\usepackage{longtable}
\usepackage[english]{babel}
\usepackage{color}

\begin{document}

\title{Nonlinear dielectric relaxation of polar liquids} 

\author{Tuhin Samanta} 
\affiliation{Department of Physics, Arizona State University,  PO Box 871504, Tempe, AZ 85287-1504}
\author{Dmitry V.\ Matyushov }
\affiliation{School of Molecular Sciences and Department of Physics, Arizona State University,  PO Box 871504, Tempe, AZ 85287-1504 }
\email{dmitrym@asu.edu}

\begin{abstract}
Molecular dynamics of two water models, SPC/E and TIP3P, at a number of temperatures is used to test the Kivelson-Madden equation connecting single-particle and collective dielectric relaxation times through the Kirkwood factor. The relation is confirmed by simulations and used to estimate the nonlinear effect of the electric field on the dielectric relaxation time. We show that the main effect of the field comes through slowing down of the single-particle rotational dynamics and the relative contribution of the field-induced alteration of the Kirkwood factor is insignificant for water. Theories of nonlinear dielectric relaxation need to mostly account for the effect of the field on rotations of a single dipole in a polar liquid.             
\end{abstract}
\maketitle

\section{Introduction}
\label{sec:1}
%

%
Rotational dynamics in liquids are affected by mutual interactions between the  molecules. One can experimentally distinguish between rotational dynamics of a single molecule and collective dynamics probed by applying a uniform perturbation to the bulk sample. The single-particle rotational dynamics are accessible by NMR\cite{Qvist:2009kx} and time-resolved IR\cite{Bakker:2010ji} spectroscopies and by incoherent neutron scattering.\cite{BeeBook} The collective rotational dynamics of molecular dipoles is reported by dielectric spectroscopy.\cite{Bottcher:78} 

The rotational relaxation time of a single dipole in the liquid is associated with the time autocorrelation function of the molecular dipole moment $\bm{\mu}(t)$. By defining the unit vector specifying the dipole orientation $\hat{\mathbf{u}}(t)=\bm{\mu}(t)/\mu$, one obtains
\begin{equation}
\phi_s(t)=\langle \hat{\mathbf{u}}(t)\cdot\hat{\mathbf{u}}\rangle ,
\label{eq10}     
\end{equation}
where the angular brackets denote an equilibrium ensemble average and we use the notation $\hat{\mathbf{u}}=\hat{\mathbf{u}}(0)$. The integral single-particle relaxation time follows as the time integral of $\phi_s(t)$
\begin{equation}
\tau_s =\int_0^\infty dt \phi_s(t) .   
\label{eq11}
\end{equation}

The collective rotational dynamics is defined by the dynamics of the macroscopic dipole moment of the sample $\mathbf{M}(t) = \sum_j \bm{\mu}_j(t)$, where the sum $j=1,\dots,N$ runs over all $N$ dipole moments in the sample.  The corresponding normalized collective time correlation function is
\begin{equation}
\phi_M(t) = \left[\langle \delta \mathbf{M}^2\rangle \right]^{-1} \langle \delta \mathbf{M}(t)\cdot \delta \mathbf{M} \rangle,  
\label{eq12}
\end{equation}
where $\delta\mathbf{M}(t) =\mathbf{M}(t) -\langle \mathbf{M}\rangle$ and $\langle \mathbf{M}\rangle=0$ in an isotropic sample. The distinction between $\phi_s(t)$ and $\phi_M(t)$ arises from time-dependent cross-correlations\cite{Gabriel:2020kc,Pabst:2021cc} between a given target dipole $\bm{\mu}_1(0)$ chosen at $t=0$ with the rest of the dipoles in the liquid at a later time $t$ 
\begin{equation}
\phi_c(t) = \sum_{k=2}^N \langle \hat{\mathbf{u}}_k(t)\cdot \hat{\mathbf{u}}_1\rangle .  
\label{eq13}  
\end{equation}
In contrast to $\phi_s(t)$ and $\phi_M(t)$, which are both normalized to unity at $t=0$, the cross correlation function at $t=0$ yields the deviation of the Kirkwood factor from the limit of uncorrelated dipoles ($g_\text{K}=1$): $\phi_c(0)=g_\text{K}-1$,
\begin{equation}
g_K = \mu^{-1} \langle \hat{\mathbf{u}}_1\cdot \mathbf{M}\rangle .
\label{42}  
\end{equation}

From definitions in Eqs.\ \eqref{eq10}, \eqref{eq12}, and \eqref{eq13}, it is easy to see that the collective and single-particle time correlation functions are related by the following equation
\begin{equation}
g_\text{K} \phi_M(t) = \phi_s(t) + \phi_c(t) .  
\label{eq14}
\end{equation}
One can further define the collective relaxation time $\tau_M$ by replacing $\phi_s(t)$ with $\phi_M(t)$ in Eq.\ \eqref{eq11}. This integral definition for the relaxation times yields the following relation 
\begin{equation}
g_\text{K} \tau_M = \tau_s + (g_\text{K}-1)\tau_c,   
\label{eq15}
\end{equation}
where $\tau_c$ is the relaxation time defined as the time integral of the normalized $\phi_c(t)$
\begin{equation}
\tau_c = \int_0^\infty dt \phi_c(t)/\phi_c(0) .
\label{eq16}  
\end{equation}

Kivelson and Madden\cite{Kivelson:1975eg,Madden:84} suggested a simple relationship between the single-particle and collective relaxation times 
\begin{equation}
\tau_M = \tau_s g_\text{K} . 
\label{eq17}  
\end{equation}
In contrast to standard expectations anticipating $\tau_M>\tau_s$, this equation allows both slowing down and speedup of collective dynamics compared to the single-particle dynamics. Given the exact relation between three relaxation times in Eq.\ \eqref{eq15}, Eq.\ \eqref{eq17} yields a nontrivial result for the relaxation time of cross-correlations between the dipoles in the liquid
\begin{equation}
\tau_c = \tau_s \left(1+ g_\text{K}\right), 
\label{eq18}  
\end{equation}
which applies assuming $g_\text{K} \ne 1$. This equation states that out of three time scales characterizing the dynamics of polar liquids, $\tau_s$, $\tau_M$, and $\tau_c$, the relaxation of cross-correlations of the liquid dipoles is the slowest process. It can potentially be observed as a separate Debye peak in the dielectric relaxation spectrum.\cite{Pabst:2020,Pabst:2021cc} One, nevertheless, has to keep in mind that the relative weights of the self and cross correlation functions in the overall dielectric function $\epsilon(\omega)$ are set by the dynamic Kirkwood-Onsager equation. For strongly polar liquids, it can be written in the form of the Debye equation\cite{DMbook} 
\begin{equation}
\epsilon(\omega) -\epsilon_\infty = \Delta \epsilon\left[1+i\omega \tilde \phi_M(\omega) \right],
\label{19}
\end{equation}
where $\Delta \epsilon = \epsilon_s -\epsilon_\infty$ is the increment of the static dielectric constant $\epsilon_s$ over the high-frequency limit $\epsilon_\infty$ and $\tilde\phi_M(\omega)$ is the Fourier-Laplace transform\cite{Hansen:13} of the time correlation function $\phi_M(t)$ in Eq.\ \eqref{eq12}. 

Equation \eqref{19} can be rewritten in a more compact form as
\begin{equation}
\epsilon(\omega)-\epsilon_s = i\omega \Delta\epsilon \tilde\phi_M(\omega) = i\omega \frac{\Delta\epsilon}{g_K}\left[\tilde\phi_s(\omega)+\tilde\phi_c(\omega) \right], 
\label{191} 
\end{equation}
where Eq.\ \eqref{eq14} was used in the second step. From this equation, the ratio of amplitudes of the cross-correlation and self relaxation processes in the dielectric spectrum is equal to $g_K-1$, independently of the corresponding relation between the relaxation times. A simplistic separation of the dielectric spectrum into the self and cross-correlation components\cite{Pabst:2021cc} does not apply when this exactly prescribed ratio of line amplitudes is not satisfied. Therefore, if cross-correlations account for the appearance of high-intensity, low-frequency Debye peaks in the polarization dynamics of low-temperature liquids, they have to be assigned to some sub-sets of cross-correlations. This assignment would also imply that other subsets produce negative cross-correlations to account for the entire relative weight of $g_K-1$ in the sum rule.  

The Kivelson-Madden equation is derived from Mori-Zwanzig projection operators formalism\cite{Mori:1965} and is based on defining $\hat{\mathbf{u}}(t)$, $\partial_t\hat{\mathbf{u}}(t)$, $\mathbf{M}(t)$, and $\partial_t\mathbf{M}(t)$ as a set of slow dynamic variables for which memory equations are established. The restriction to a reduced set of variables is valid when $\tau_s^2k_\text{B}T/I\gg 1$, where $I$ is the molecular moment of inertia. This parameter is $\simeq 300$ for water molecules at $T\simeq 300$ K and the restricted dynamical subspace is justified. The complete solution of the theory is 
\begin{equation}
     \tau_M = \tau_s g_K( 1 + Nh),   
     \label{192}
\end{equation}
where $h$ is given in terms of cross-correlations of orthogonally (anomalously\cite{BalucaniBook}) propagated angular accelerations, $\partial_t^2\hat{\mathbf{u}}_1$ and $\partial_t^2\hat{\mathbf{u}}_2$, of distinct molecular dipoles (Eq.\ (B10) in Ref.\ \onlinecite{Kivelson:1975eg}). One arrives at Eq.\ \eqref{eq17} if these cross-correlations are neglected. The physical meaning of the Kivelson-Madden prescription is that it relates the alteration of single-molecule dynamics due to many-body interactions in the liquids solely to static correlations of dipolar orientations at $t=0$. 

\begin{figure}
\includegraphics*[clip=true,trim= 0cm 1.5cm 0cm 0cm,width=9cm]{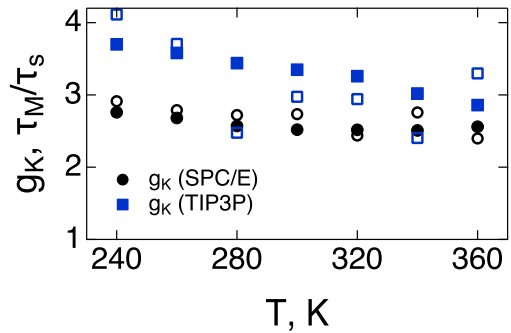}
\caption{$g_\text{K}$ (filled points) and $\tau_M/\tau_s$ (open points) for TIP3P (squares) and SPC/E (circles) water vs $T$. Points are results of MD simulations. }
\label{fig:1}  
\end{figure}

The derivation of the Kivelson-Madden relation involves some approximations, particularly in its simplified form in Eq.\ \eqref{eq17}, and it has remained a conjecture for many years since its introduction.\cite{Kivelson:1975eg} Nevertheless, recent experimental studies by Weing{\"a}rtner and co-workers\cite{Volmari:2002ti,Weingartner:2004} and molecular dynamics (MD) simulations by Steinhauser and co-workers\cite{Braun:2014fs,Honeger:2018} have produced  evidence of its accuracy. Here, we  use classical molecular dynamics (MD) simulations of force-field water to provide additional tests and to use this result toward the goal of modeling the nonlinear retardation of polar dynamics by the applied electric field. Since this relation does not specify temperature and should be valid at least in some range of temperatures, we use temperature as an additional variable to alter all three parameters in this equation. The relaxation times $\tau_a$, $a=M,s$ and $g_\text{K}$ are calculated from configurations produced by classical MD  simulations of SPC/E\cite{Berendsen:87} and TIP3P\cite{tip3p:83} water models at different temperatures. Figure \ref{fig:1} shows the result of these calculations supporting Eq.\ \eqref{eq17} within simulation uncertainties. This result is next applied to estimate the nonlinear alteration of the collective dielectric dynamics with the applied external field.    

\section{Nonlinear dynamics}
Rotational dynamics of liquid dipoles is a liquid's intrinsic property,  independent of the applied external field in the linear response approximation.\cite{Hansen:13}  As the strength of the external field increases, nonlinear effects start to affect dynamics and relaxation times shift with increasing field strength: $\tau_a^E=\tau_a(E)$, $a=M,s,c$. Effects of the field on the relaxation times are difficult to predict beyond single-particle dynamics.\cite{Dejardin:2000aa} Nevertheless, Eq.\ \eqref{eq17} offers a convenient solution
\begin{equation}
d\tau_M^E/df_E = d(\tau_s^E g^E_\text{K})/df_E,  
\label{eq20}
\end{equation}
where the derivative is taken at zero field $E=f_E=0$. 

We use the free energy of polarizing the dielectric sample per molecule of the sample to quantify the field strength 
\begin{equation}
f_E = \frac{\beta\epsilon}{8\pi\rho} E^2  ,
\label{eq21}
\end{equation}
where $\rho=N/V$, $V$ is the sample volume, and $\beta=(k_\text{B}T)^{-1}$ is the inverse temperature. Further, $\epsilon$ is the linear dielectric constant of the material, i.e., the dielectric constant in the limit $E=0$.  The parameter $f_E$ is a natural scale for gauging the field strength comparing the polarization energy to thermal energy at the scale of a single molecule.\cite{DMjcp3:15} It amounts to $f_E\simeq 2\times 10^{-4}$ for water placed in the field of $E=10^5$ V/cm often employed in experiment.\cite{Richert:2014wa} This estimate indicates that most experiments do not produce polarization energies significantly affecting molecular motion. Therefore, only small deviations from linear static and dynamic properties of dielectrics can be achieved. Measurable deviations from linearity scale with $E^2\propto f_E$ in the lowest non-vanishing order in the field.\cite{Richert:2014wa} 

The effect of the field on the dielectric relaxation time come through changes in the single-particle relaxation time and the Kirkwood factor representing the collective effects of statistical correlations between the dipoles (Eq.\ \eqref{42}). Direct calculations of changes in the dynamics in the applied electric field is not easy to perform by simulations since very high fields, significantly perturbing the orientational liquid structure,\cite{Yeh:1999hb} are required to accumulate sufficient statistics. We, therefore, use an alternative approach allowed by the linear response approximation.\cite{Hansen:13} The single-particle correlation function is altered by the field of external charges (the vacuum field) $E_0$ to become $\phi_s^E(t)$ and we use perturbation theory to find the change $\Delta \phi_s^E(t)=\phi_s^E(t)-\phi_s(t)$ in the lowest order in $f_E$.    

The direct application of the perturbation theory leads to a change in the correlation function $\Delta \phi_s^E(t)$ quadratic in $E_0$ in the lowest order of the perturbation theory. A full solution of the problem requires accounting for quadratic field effects on the Liouville dynamics,\cite{Kubo:59} which are difficult to achieve in analytical techniques. A simplification is possible if the dynamics are fast and the field effect is mostly accounted for through the alteration of the initial, $t=0$, dipolar orientations in the time correlation function. This approach follows the philosophy leading to the Kivelson-Madden relation (Eq.\ \eqref{eq17}) and should be equally applicable if this basic prescription holds. 

The field of external charges $\mathbf{E}_0$ perturbs the system Hamiltonian from the unperturbed function $H_0$ to $H'=H_0- \mathbf{M}\cdot \mathbf{E}_0$. The perturbation of the statistics of the initial orientations $\hat{\mathbf{u}}=\hat{\mathbf{u}}(0)$  is given by a series in even powers of $E_0$. In contrast, dielectric spectra are recorded in terms of the uniform Maxwell field $E$, which, in the plane capacitor, is equal to the voltage at the capacitor plates divided by their separation. In order to express the solution in terms of $E$, we consider a slab sample and direct the external field first along the $z$-axis perpendicular to the plates, $E_{0z}=\epsilon E$, followed by directing the field along the $x$-axis in the capacitor's plane, $E_{0x}=E$.\cite{Jackson:99} Combining the linear in $E_0^2$ response along the $z$-axis with two equal responses along the $x$-axis and $y$-axis, one gets  $\Delta \phi_s^E(t)$ proportional to $E^2$. The final result can be conveniently re-written in terms of $f_E$ in Eq.\ \eqref{eq21} as follows  
\begin{equation}
\frac{\Delta \phi_s^E(t)}{\phi_s(t)} = 9yf_E \frac{\epsilon}{2\epsilon^2+1}\,\Psi(t).   
\label{eq22}
\end{equation}
The nonlinear time correlation function is   
\begin{equation}
 \Psi(t) = \phi_s(t)^{-1} \langle \hat{\mathbf{u}}(t)\cdot\hat{\mathbf{u}}(0)\delta [\mathbf{M}^2]\rangle ,
 \label{eq23}
\end{equation}
where $\delta [\mathbf{M}^2]=(\mathbf{M}^2 - \langle\mathbf{M}^2\rangle)/\mu^2=N\delta g_K$ specifies the fluctuation of the Kirkwood factor scaled with the number of liquid molecules $N$. This scaling suggest that the correlation $\langle \hat{\mathbf{u}}(t)\cdot\hat{\mathbf{u}}(0)\delta g_K\rangle$ scales as $N^{-1}$ to allow a finite value in the thermodynamic limit. Further, the parameter $y=(4\pi/9)\beta \mu^2\rho$ in Eq.\ \eqref{eq22} is the standard dipolar density parameter of the dielectric theories.\cite{Boettcher:73} The time correlation function $\Psi(t)$ satisfies the boundary conditions $\Psi(0)=\dot \Psi(0)=0$. It implies that $\Psi(t)\propto t^2$ at low $t$. On the other hand, one expects a linear time dependence at intermediate times. This result is derived from the following empirical arguments.

\begin{figure}
\includegraphics*[clip=true,trim= 0cm 1cm 0cm 0cm,width=9cm]{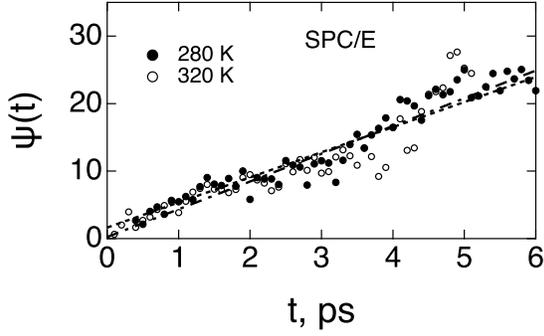}
\caption{$\Psi^E(t)$ vs $t$ for SPC/E water from simulations at 280 K (filled circles) and 320 K (open circles). The dashed and dash-dotted lines are linear fits through the points at 280 K and 320 K, respectively.    }
\label{fig:2}  
\end{figure}

Assume that the time correlation function $\phi_s^E(t)$ is given by an exponential decay with the decay exponent altered from the no-field relaxation time $\tau_s$ to the in-field relaxation time $\tau_s^E$ according to the empirical relation\cite{Richert:2014wa} anticipating linear scaling with $E^2$ 
\begin{equation}
\tau_s^E=\tau_s(1 + a_\tau f_E).  
\label{eq24}
\end{equation}
This relation implies
\begin{equation}
\ln[ \phi_s^E(t)/\phi_s(t)] \simeq \Delta \phi_s^E(t)/\phi_s(t)\simeq (t/\tau_s) a_\tau f_E .    
\label{eq25}
\end{equation}
Combining this relation with Eqs.\ \eqref{eq22} and \eqref{eq23}, one arrives at the relation between the coefficient of dynamical slowing down of the single-particle dynamics and the nonlinear time correlation function accessible from simulations 
\begin{equation}
a_\tau =   \frac{9y\epsilon\tau_s}{2\epsilon^2+1} \frac{\Delta\Psi(t)}{\Delta t} ,
\label{eq26}
\end{equation}
where $\Delta\Psi(t)/\Delta t$ specifies the slope of the linear portion of $\Psi(t)$ (dashed line in Fig.\ \ref{fig:2}).  From this equation, the derivative of the collective relaxation time over $f_E$ becomes
\begin{equation}
  d\tau_M^E/df_E=\tau_s\left( a_\tau g_K + \partial g_K^E/\partial f_E\right). 
  \label{26}
\end{equation}

\section{Results of simulations}
Molecular dynamics simulations of SPC/E and TIP3P water models were carried out as explained in the supplementary material. Figure \ref{fig:1} shows the temperature dependent Kirkwood factors calculated from simulations of two water models (filled points). The standard route to the dielectric constants $\epsilon$ is through computing the variance of the dipole moment of the cubic simulation cell with the volume $V$. When tin-foil boundary conditions are implemented in the Ewald sum protocol for the electrostatic interactions,\cite{Neumann:86}  one obtains
\begin{equation}
\epsilon = 1 + \frac{4\pi\beta}{3V} \langle \delta \mathbf{M}^2\rangle .
\label{eq27}  
\end{equation}
The Kirkwood factor then follows from the Kirkwood-Onsager equation\cite{Frohlich} 
\begin{equation}
(\epsilon-1)(2\epsilon+1) = 9yg_\text{K}\epsilon .   
\label{eq28}
\end{equation}
This approach does not address the issue of the effect of finite size of the simulation box on the computed values and an alternative approach was used here. 

\begin{figure}
\includegraphics*[clip=true,trim= 0cm 1.5cm 0cm 0cm,width=9cm]{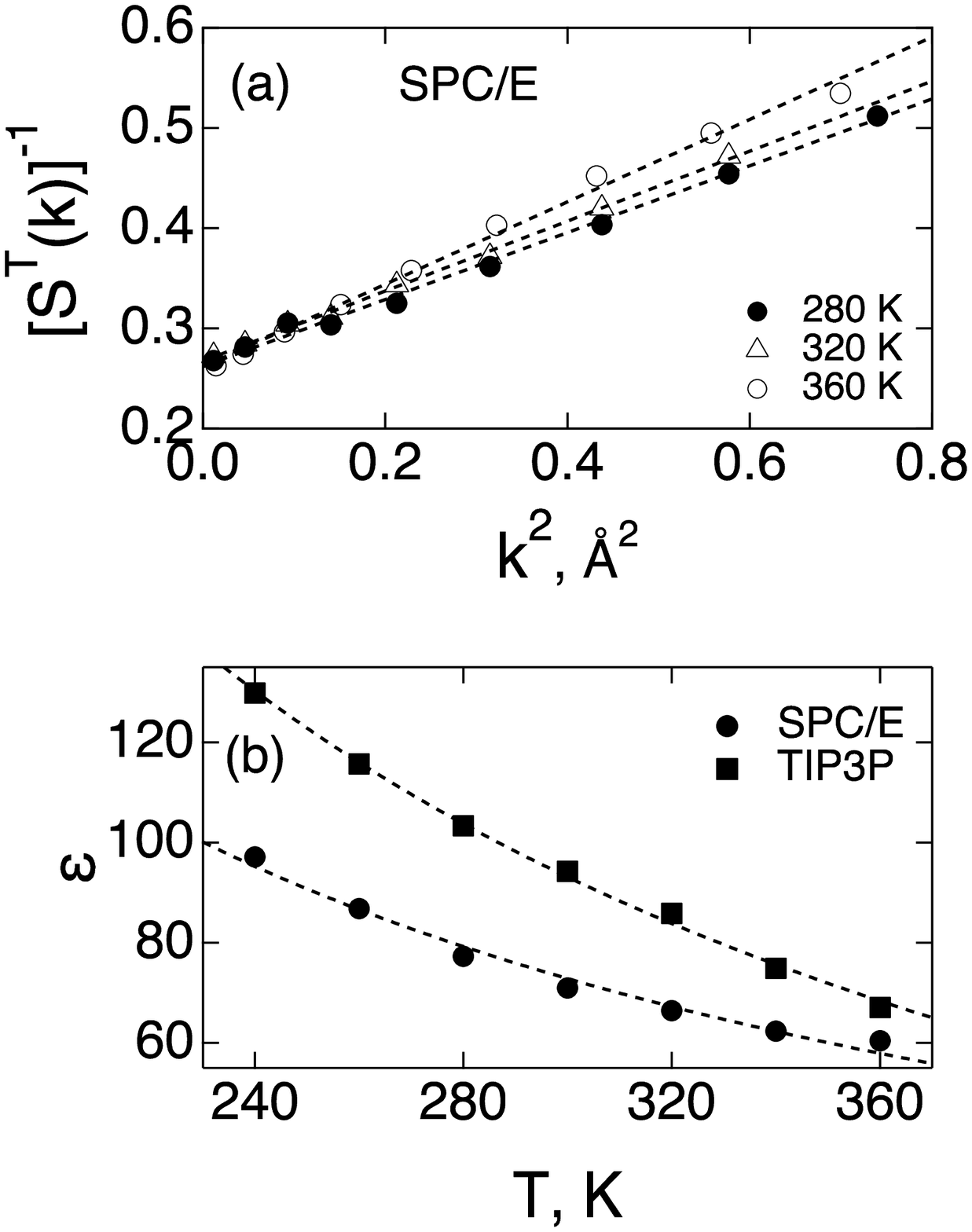}
\caption{(a) $[S^T(k)]^{-1}$ vs $k^2$ for SPC/E water at different temperatures. The dashed lines are linear fits through the points calculated from MD trajectories. (b) $\epsilon(T)$ obtained from combining Eq.\ \eqref{30}. The dashed lines are fits to the function $a+b/T$ with $a=-55.42$ (TIP3P), $-16.73$ (SPC/E) and $b=44565$ K (TIP3P), 26870 K (SPC/E).  }
\label{fig:3}  
\end{figure}

The calculation of $\epsilon$ and $g_K$ was based here on computing the transverse dipolar structure  factor\cite{Fonseca:90,DMjcp1:04}
\begin{equation}
S^T(k) = \frac{3}{2N}\left\langle\sum_{i,j} \left[(\hat{\mathbf{u}}_i\cdot \hat{\mathbf{u}}_j)- (\hat{\mathbf{u}}_i\cdot \hat{\mathbf{k}})(\hat{\mathbf{k}}\cdot\hat{\mathbf{u}}_j) \right] e^{i\mathbf{k}\cdot\mathbf{r}_{ij}}  \right\rangle ,
\label{29}
\end{equation}
where $\hat{\mathbf{k}}=\mathbf{k}/k$ is the unit vector of the wavevector $\mathbf{k}=(2\pi/L)(n,l,m)$ calculated on the cubic lattice with the side length $L$; $\hat{\mathbf{u}}_i$ are orientational unit vectors of molecular dipoles and $\mathbf{r}_{ij}$ are the distances between the center of mass coordinates of molecules $i,j=1,\dots,N$. 

The macroscopic dielectric constant was calculated from the transverse structure factor by linearly extrapolating $[S^T(k)]^{-1}=[S^T(0)]^{-1}+ \Lambda^2 k^2$ from finite lattice vectors to $k=0$. A linear scaling of $[S^T(k)]^{-1}$ with $k^2$ is expected from general properties of the Ornstein-Zernike equation and the zero-value transverse structure factor is related to the dielectric constant as\cite{Wertheim:71}  
\begin{equation}
S^T(0) = (\epsilon-1)/(3y).   
\label{30}
\end{equation}
Examples of extrapolations at different temperatures are shown in Fig.\ \ref{fig:3}a. The results for $\epsilon(T)$ are presented in Fig.\ \ref{fig:3}b and are listed in Table S2 in supplementary material. The temperature slope of $\epsilon(T)$ in our calculations exceeds that from the results of Fennell et al\cite{Fennell:2012ee} (Fig.\ S4).

\begin{figure}
\includegraphics*[clip=true,trim= 0cm 1cm 0cm 0cm,width=9cm]{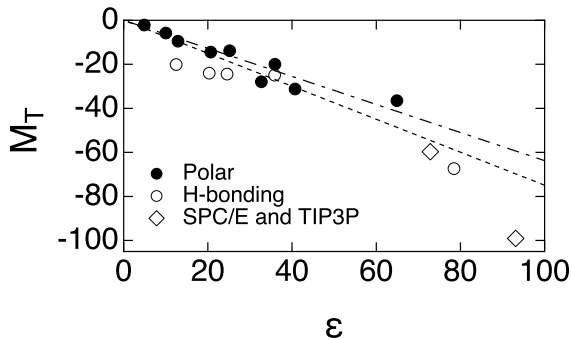}
\caption{$M_T$ calculated from Eq.\ \eqref{eq:6} for polar liquids (filled points) and alcohols and water (open points) at $T=298$ K. The experimental data are taken from Ref.\ \onlinecite{Marcus15} and the dash-dotted (polar liquids) and dashed (H-bonding liquids) lines are linear fits through the points. Two points  (open diamonds) for SPC/E ($\epsilon=71$) and TIP3P ($\epsilon=94$) water are produced with the use of MD results from Fig.\ \ref{fig:3}. }
\label{fig:4}  
\end{figure}

The temperature derivative of the dielectric constant also represents nontrivial multiparticle orientational correlations in the liquid.\cite{DMjcp1:16} The logarithmic temperature derivative of the dielectric constant becomes\cite{DMNlinDiel} (see supplementary material) 
\begin{equation}
\left(\frac{\partial \ln \epsilon}{\partial \ln T}\right)_V = \frac{3\epsilon}{2\epsilon^2 + 1} M_T .
\label{eq:6}  
\end{equation}
In this equation,
\begin{equation}
M_T = \frac{4\pi\beta}{3V} \langle (\delta\mathbf{M})^2 (\beta\delta H_0-1)\rangle 
\label{eq:7}  
\end{equation}
correlates fluctuations of the squared dipole moment with the fluctuation of the Hamiltonian $H_0$ of the unperturbed polar liquid. The parameter $M_T$  incorporates three- and four-particle orientational correlations of molecular dipoles.\cite{DMjcp1:16}  From the present MD simulations, $M_T$ is equal to $-59.7$ for SPC/E water and $-99.0$ for TIP3P water at $T=298$ K (open diamonds in Fig.\ \ref{fig:4}). Figure \ref{fig:4} compares these results to experimental values\cite{Marcus15} for polar and hydrogen-bonding liquids. The overall consistency with the experimental results for other polar liquids and with laboratory water (open circle at $\epsilon\simeq 78$) testify to the accuracy of our calculations presented in Fig.\ \ref{fig:3}. 

The Kirkwood factor $g_K(T)$ is a decaying function of temperature (Table S2). This outcome violates the prediction of the fluctuation-dissipation theorem (FDT)\cite{Kubo:66} stipulating that the variance of a macroscopic variable scales linearly with $T$. In the case of the Kirkwood factor, FDT requires $g_K\propto \langle \delta\mathbf{M}^2\rangle \propto T$, which is obviously violated by the results of both experiment\cite{DMjcp1:16} and of the present simulations.

Turning to single-molecule and collective dynamics, configurations of water produced by MD simulations were used to calculate the single-particle, $\phi_s(t)$, and collective, $\phi_M(t)$, relaxation functions. The time correlation functions were fitted to sums of two exponential functions, with one exponential mostly used for $\phi_M(t)$.\cite{Braun:2014fs} The single-molecule correlation function $\phi_s(t)$ is a sum of $\sim20$\% amplitude decay with a temperature-independent relaxation time $\simeq 0.14-0.19$ ps, followed with a slower exponential decay with the relaxation times strongly affected by temperature (Table S1). The relaxation times for the sample dipole moment, $\tau_M$, and of single-molecule rotational dynamics, $\tau_s$, were calculated separately and the ratio $\tau_M/\tau_s$ was produced. The results are shown by open points in Fig.\ \ref{fig:1}, which are compared to $g_\text{K}$ values (filled points). We indeed find a reasonable agreement with Eq.\ \eqref{eq17} within simulation uncertainties. Note that there is no need to invoke the dynamical Kirkwood correlation factor $J_K$ in Eq.\ , in contrast to $J_K=(1+Nh)^{-1}\simeq 1.7-1.8$ (Eq.\ \eqref{192}) found in Ref.\ \onlinecite{Braun:2014fs}. The ratio $\tau_s/\tau_M$ from the Kievelson-Madden equation is also much higher than the value $\simeq 64(3y/\pi)^{5/2}\exp[-6y]\simeq 3\times 10^{-13}$ at $y\simeq 6.2$ for SPC/E water proposed by D\'ejardin et al.\cite{Dejardin:2019ba}  
 
The alteration of single-molecule dynamics with the applied external field needs to be compared with the corresponding change in the Kirkwood factor to estimate two different contributions to Eq.\ \eqref{26}. The field-dependent Kirkwood factor can be defined in terms of the variance of the macroscopic dipole moment of the sample in the presence of the field
\begin{equation}
g_K^E = \frac{1}{\mu^2 N} \langle (\delta \mathbf{M})^2\rangle_E ,  
\label{31}
\end{equation}
where $\langle\dots\rangle_E$ is an ensemble average in the presence of the applied field and $\delta\mathbf{M}=\mathbf{M}-\langle \mathbf{M}\rangle_E$. By applying the perturbation expansion in terms of the external field, one obtains\cite{DMjcp3:15} (see supplementary material for derivation) 
\begin{equation}
g_K^E = g_K - \frac{(\epsilon-1)^2}{y\epsilon} f_E B_V  .
\label{32}
\end{equation}
In this equation, the parameter  
\begin{equation}
B_V=N\left[1-\frac{\langle M^4\rangle}{3\langle M^2\rangle^2} \right].  
\label{33}
\end{equation}
describes the deviation of the statistics of the macroscopic dipole moment $M$ projected on the direction of the external field from the Gaussian statistics stipulated by the central limit theorem. 

The term in the brackets in Eq.\ \eqref{33} goes to zero as $N^{-1}$. Since this term is multiplied with $N$, the parameter $B_V$ specifies the first-order correction to the Gaussian statistics of $M$ for a macroscopic sample.  From Eq.\ \eqref{32}, one obtains for the derivative of the Kirkwood factor
\begin{equation}
\frac{\partial g^E_\text{K}}{\partial f_E} = -\frac{\Delta\epsilon^2}{y\epsilon} B_V , 
\label{eq29}  
\end{equation}
where $\Delta \epsilon=\epsilon-1$ is the increment of the dielectric constant over its high-frequency limit for a nonpolarizable liquid. 

The parameter $B_V$ can be connected to the alteration of the dielectric constant with the field\cite{Chelkowski:80} $\Delta \epsilon_E=\epsilon(E)-\epsilon\propto f_E$. The relation is given in terms of the third-order dielectric susceptibility $\chi_3$, which is the first nonlinear correction to the linear susceptibility $\chi$, $\epsilon=1+4\pi\chi$
\begin{equation}
\langle P\rangle_E = \chi E + \chi_3 E^3 .  
\label{34}
\end{equation}
The relation between $\Delta \epsilon_E$ and $\chi_3$ depends, however, on the experimental setup.\cite{DMjpcm1:21} If a small sinusoidal field is combined with a constant large-amplitude bias, one gets

\begin{equation}
\Delta \epsilon_E= 12\pi \chi_3 E^2 . 
\label{35} 
\end{equation}
Alternatively, when a large-amplitude oscillating field with zero bias is used, one finds \cite{DMjpcm1:21}
\begin{equation}
  \Delta \epsilon_E = 3 \pi \chi_3 E^2 . 
  \label{36}
\end{equation}
By applying this last relation and the connection between $\Delta\epsilon_E$ and $\chi_3$ from the perturbation expansion\cite{DMjcp3:15,DMNlinDiel} (also see the supplementary material) one obtains
\begin{equation}
\frac{\partial \Delta\epsilon_E}{\partial f_E} = - \frac{3}{4} \epsilon\Delta\epsilon^2 B_V . 
\label{37} 
\end{equation}
Combining Eqs.\ \eqref{eq29} and \eqref{37}, one finally obtains
\begin{equation}
   \frac{\partial g^E_\text{K}}{\partial f_E} = \frac{4}{3y\epsilon^2} \frac{\partial \Delta\epsilon_E}{\partial f_E} . 
   \label{38}
\end{equation}
This equation is an exact result limited only by truncation of the higher-order expansion terms. Considering small deviations of the Kirkwood factor and the dielectric constant from the zero-field values, one can apply the Kirkwood-Onsager equation to write
\begin{equation}
\Delta \ln g_K^E = \frac{12}{(\epsilon-1)(2\epsilon+1)} \Delta\epsilon_E .  
\label{41} 
\end{equation}

Linear scaling of $\Delta\epsilon_E$ with $E^2$ is often characterized with the empirical Piekara factor\cite{Piekara:62,Davies:1978tw,Richert:2014wa}  $a=\Delta\epsilon_E/E^2$. Equation \eqref{38} thus provides the link between the field-induced change in the Kirkwood factor and the Piekara factor
\begin{equation}
 \Delta g_K^E = \frac{4a}{3y\epsilon^2} E^2 . 
 \label{39} 
\end{equation}
The Piekara factor quantifies the field-induced alteration of average cosines between the dipoles in the liquid ($g_K$, Eq.\ \eqref{39}) or the extent of the non-Gaussian statistics of the macroscopic dipole moment of the sample (Eq.\ \eqref{37}).  

The experimental slope for water\cite{Davies:1978tw} (at constant pressure) is $(\partial \Delta\epsilon_E/\partial (E^2))_P\simeq -0.8\times 10^{-15}$ $\mathrm{m^2/V^2}$, which yields $\partial \Delta\epsilon_E/\partial f_E\simeq -3.2\times 10^2$ and $\partial g_K^E/\partial f_E\simeq -0.01$ for ambient water. A similar estimate based on simulation data\cite{2016PhRvB..93n4201Z} for SPC/E water yields $(\partial \Delta\epsilon_E/\partial (E^2))_V\simeq -0.16\times 10^{-15}$ $\mathrm{m^2/V^2}$ and $(\partial g^E_\text{K}/\partial f_E)_V \simeq -0.003$ for SPC/E water at 300 K.    

The correlation function $\Psi(t)$ involves many-particle dynamical and static correlations between dipoles in the liquid. Its calculation requires long trajectories to achieve convergence. Nevertheless, the advantage of this protocol is the ability to use MD configurations in the absence of the field and thus produce the retardation factor in the limit of a weak field typical for experimental conditions. 

Figure \ref{fig:2} illustrates $\Psi(t)$ for SPC/E water at two temperatures. The range of times was limited given that we expect linear scaling to hold for times comparable to the single-molecule rotational time $\tau_s$. The values of $a_\tau$ (Eq.\ \eqref{eq26}) obtained from the slopes are $\simeq 17.1$ at $T=280$ K and $\simeq 8.6$ at $T=320$ K. The linear slope of $\Psi(t)$ is higher at 320 K compared to 280 K, but it is compensated by a lower $y\tau_s$ in Eq.\ \eqref{eq26}. The convergence of $\Psi(t)$ is poor and only qualitative conclusions can be made. Nevertheless, the values of $a_\tau$ from the linear slopes significantly exceed the contribution to $d\tau_M^E/df_E$ from the field dependence of the Kirkwood factor (the second term in Eq.\ \eqref{26}). This allows one to relate the field-induced alteration of the collective relaxation time to the change in the single-molecule dynamics
\begin{equation}
\Delta \ln \tau_M^E \simeq a_\tau f_E.  
\label{40}
\end{equation}
Converting $f_E$ to $E^2$, one obtains for SPC/E water at $E=100$ kV/cm typically used in nonlinear dielectric experiments\cite{Richert:2014wa} $\Delta \ln \tau_M^E$ amounting to 0.2-0.4\%. The reported values of changes in the Debye relaxation time are within 0.14--1.65 \% at this field magnitude.\cite{RichertBook:2018}    

The coefficient $a_\tau$ in Eq.\ \eqref{eq24} is positive in our calculations. An applied electric field, therefore, slows the single-molecule dynamics down. In contrast, the Kirkwood factor is lowered by the field (Eq.\ \eqref{41}) and this term in Eq.\ \eqref{26} speeds the dynamics up. The single-molecule and collective aspects of the nonlinear dynamics thus oppose each other. However, the effect of the field on the single-molecule dynamics is the dominant contribution to the alteration of the relaxation time. This is a natural result in the Kivelson-Madden framework given small values for  the variation of the Kirkwood factor with the applied field in Eq.\ \eqref{26}. 

Relaxation of the dipole moment of water is mostly single-exponential. The present simulations, therefore, do not address the nonlinear dielectric dynamics of low-temperature polar liquids characterized by dispersive (stretched) relaxation functions.\cite{Richert:2014wa}   Stretching exponents increase approaching unity for more polar liquids thus making them closer to canonical Debye polar fluids.\cite{Paluch:physrevlett.116.025702} The work of Keyes and Kivelson,\cite{KeyesKivelson} preceding the Kivelson-Madden development, had suggested that single-particle and collective correlation functions carry similar mathematical forms. Based on this prediction, $\tilde\phi_s$ and $\tilde\phi_c$ in Eq.\ \eqref{191} are expected to be stretched to a comparable degree. It remains to be seen whether the Kivelson-Madden equation in its simplified form (Eq.\ \eqref{eq17}) will hold for such more complex polarization dynamics.

\section{Conclusions}
The Kivelson-Madden equation connects the collective and single-molecule dynamics of polar liquids through the Kirkwood factor (Eq.\ \eqref{42}) responsible for statistical correlations between the dipoles in the liquid. The equation, therefore, views the collective dynamics as the dynamics of individual dipoles corrected for static correlations between them. We have proved the equation to hold, within simulation uncertainties, for SPC/E and TIP3P water models in the 240--360 K range of temperatures. It was further used to estimate the nonlinear effect of the external field on the collective dielectric dynamics reported by nonlinear dielectric spectroscopy. The variation of the single-molecule dynamics of the liquid dipoles is shown to be the dominant effect in the dielectric slowing down. This result significantly simplifies the development of formal theories of nonlinear dielectric relaxation since only the effect of the field on the dynamics of a single dipole needs to be accounted for.      

\section*{Supplementary material}
See supplementary material for the simulation protocols, data analysis, and derivation of equations presented in the text.

\acknowledgments 
This research was supported by the National Science Foundation (CHE-2154465). CPU time was provided by the National Science Foundation through XSEDE resources (TG-MCB080071) and through ASU's Research Computing.   

\section*{DATA AVAILABILITY}
 
The data that supports the findings of this study are available within the article and its supplementary material.

\bibliography{chem_abbr,dielectric,dm,statmech,protein,liquids,solvation,dynamics,simulations,surface,water,glass,nano}

\end{document}